\documentclass[aps,twocolumn,pra,superscriptaddress,showpacs,tightenlines]{revtex4}
\usepackage{amssymb}
\usepackage{amsmath}
\usepackage{graphicx}
\usepackage{epsfig}
\usepackage{subfigure}
\usepackage{amsfonts}

\begin{document}
\title{Quantum switch for single-photon transport in a coupled superconducting transmission line resonator array}
\author{Jie-Qiao Liao}

\affiliation{Institute of Theoretical Physics, Chinese Academy of
Sciences, Beijing, 100190, China}
\author{Jin-Feng Huang}
\affiliation{Key Laboratory of Low-Dimensional Quantum Structures
and Quantum Control of Ministry of Education, and Department of
Physics, Hunan Normal University, Changsha 410081, China}
\author{Yu-xi Liu}
\affiliation{Institute of Microelectronics, Tsinghua University,
Beijing 100084, China}
\author{Le-Man Kuang}
\affiliation{Key Laboratory of Low-Dimensional Quantum Structures
and Quantum Control of Ministry of Education, and Department of
Physics, Hunan Normal University, Changsha 410081, China}
\author{C. P. Sun}
\affiliation{Institute of Theoretical Physics, Chinese Academy of
Sciences, Beijing, 100190, China}

\begin{abstract}
We propose and study an approach to realize quantum switch for
single-photon transport in a coupled superconducting transmission
line resonator (TLR) array with one controllable hopping
interaction. We find that the single-photon with arbitrary
wavevector can transport in a controllable way in this system. We
also study how to realize controllable hopping interaction between
two TLRs via a Cooper pair box (CPB).
When the frequency of the CPB is largely detuned from those of the
two TLRs, the variables of the CPB can be adiabatically eliminated
and thus a controllable interaction between two TLRs can be
obtained.
\end{abstract}
 \maketitle

\narrowtext

Coupled cavity arrays (CCAs)~\cite{CCAreview} have recently
attracted considerable attentions of both theorists and
experimentalists. The CCAs have been proposed to implement quantum
simulators for many-body physics, such as discovering new matter
phases of photons~\cite{HBP06,GTCH06,RF07} and providing a new
platform to study spin systems~\cite{ASB07,HBP07a}. The CCAs are
also suggested to manipulate photons for optical quantum
information processing~\cite{HLSS08,BAB07,ASYE07}. Moreover,
photon transport in the CCAs has been
investigated~\cite{ZLS07,ZGLSN08,HZSS07,ZGSS08,GIZS08}. There are
several possible ways to construct the CCAs, for example: (i)
coupled defect cavities in photonic crystals~\cite{Vuckovic}; (ii)
coupled toroidal microresonators~\cite{AKSV03}; and (iii) coupled
superconducting transmission line resonators
(TLRs)~\cite{ZGLSN08,HZSS07}.

In CCAs, there have been many proposals to realize quantum
switch~\cite{sun,Switch}, which is used to control single-photon
transport~\cite{ZGLSN08,Lukin-np,Fanpaper1,Fanpaper2}. For
example, the reflection and transmission of photons in a coupled
resonator waveguide can be controlled by a tunable two-level
quantum system~\cite{ZGLSN08,Switch}, acting as a controller.

Here, we study another approach to control the single-photon
transport in a CCA, which consists of a chain of
TLRs~\cite{Wal04,Blais04}. In our proposal, the controllable
transport is realized by a tunable coupling. As we know, how to
control coupling between two solid devices is a major challenge in
scalable quantum computing
circuits~\cite{you,you1,liu,miro,yingdan,yingdan2,Hu}. To obtain a
tunable coupling,  we propose that a Cooper pair box (CPB) acts as a coupler. When the frequency of
the coupler is largely detuned from those of the two resonators, the
variables of the coupler can be adiabatically eliminated and thus a
controllable interaction can be induced. Compared with previous
approach~\cite{ZGLSN08}, this approach has following advantage:
dynamical variables of the coupler are adiabatically eliminated,
therefore the coupler is a passive controlling element, which makes
robust to prevent from the environment of the coupler.

\begin{figure}[tbp]
\includegraphics[bb=100 255 496 491, width=8 cm]{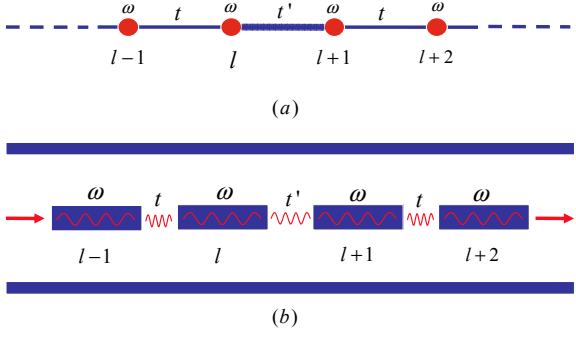}
\caption{(Color online) Schematic configuration for controllable
transport of single photon: (a) one-dimensional site lattice model
for the coupled cavity array; (b) schematic diagram of coupled
superconducting transmission line resonator
array.}\label{configuration}
\end{figure}

As shown in Fig.~\ref{configuration}, one-dimensional CCA is a chain
of $N$ cavities, each is only coupled to its nearest-neighbor ones,
Fig.~\ref{configuration}(a) and (b) are the site lattice model and
the schematic diagram of coupled TLR array, respectively. The TLRs
are assumed to have the same frequency. We also assume that the
coupling strength between two nearest-neighbor TLRs is the same,
except one between the $l$-th and $(l+1)$-th TLRs. The Hamiltonian
of the system is
\begin{eqnarray}
\label{Hamiltonian}H&=&\omega\sum_{n}a_{n}^{\dag}a_{n}-t\sum_{n}(a_{n}^{\dag}a_{n+1}+a_{n+1}^{\dag}a_{n})\nonumber\\
&&-\lambda t(a_{l}^{\dag}a_{l+1}+a_{l+1}^{\dag}a_{l}),
\end{eqnarray}
hereafter we take $\hbar=1$. Here, we assume that all TLRs have the
same frequency $\omega$. $a^{\dag}_{n}$ and $a_{n}$ are the creation
and annihilation operators of $n$-th TLR; $t$ is the coupling
strength between the $n$-th ($n\neq l$) and $(n+1)$-th TLRs;
$\lambda=(t'-t)/t$ is introduced to denote the relation between $t$
and $t'$, where $t'$ is the coupling strength between the
$l$-\textrm{th} and $(l+1)$-\textrm{th} TLRs. Obviously,
$-1<\lambda<0$ corresponds to $0<t'<t$, while $\lambda\geq 0$
implies $t'\geq t$. Below we will first study how to control the
single-photon transport by changing coupling strength $t'$, and then
answer question how to realize controllable coupling $t'$.

In the case of $t'=t$, the Hamiltonian in Eq.~(\ref{Hamiltonian})
is reduced to the usual bosonic tight binding model
$H_{\textrm{btb}}=\omega\sum_{n}a_{n}^{\dag}a_{n}-t\sum_{n}(a_{n}^{\dag}a_{n+1}+a_{n+1}^{\dag}a_{n})$
as shown in Ref.~\cite{Data}, which describes an $N$-site lattice
model with nearest-neighbor coupling. It is well known that, under
the periodic boundary condition, the bosonic tight binding
Hamiltonian can be diagonalized as
$H_{\textrm{btb}}=\sum_{k}\Omega_{k}a_{k}^{\dag}a_{k}$ by using
the Fourier transformation
$a_{k}=\sum_{n}\exp(iknd_{0})a_{n}/\sqrt{N}$, where $d_{0}$ is the
site distance. Below,  $d_{0}$ is taken as units. We choose the
wavevectors $k=2\pi m/N$ within the first Brillouin zone, i.e.,
$-N/2<m\leq N/2$. The corresponding dispersion relation is
$\Omega_{k}=\omega-2t\cos k$, which is an energy band structure.
For $t>0$, the wavevectors $k=\pm\pi/2$ correspond to the energy
band center, while the wavevectors $k=0$ and $k=\pm\pi$ correspond
to the bottom and top of the energy band, respectively.

Let us now define a total excitation number operator
$\hat{N}=\sum_{n}a_{n}^{\dag}a_{n}$. It is straightforward to show
that $\hat{N}$ commutes with the model
Hamiltonian~(\ref{Hamiltonian}), i.e., $[\hat{N},H]=0$, which
implies that the total excitation number $\hat{N}$ is a conserved
observable. We now restrict our discussion to the single excitation
subspace since we only consider the single-photon transport. In this
case, a general state can be written as
$|\Omega\rangle=\sum_{n}A_{n}|1_{n}\rangle$, where we have
introduced the basis state
$|1_{n}\rangle=|0\rangle\otimes...\otimes|1\rangle_{n}\otimes...\otimes|0\rangle$,
which represents the state that the $n$-\textrm{th} TLR has one
photon while other TLRs have no photon. $A_{n}$ is the probability
amplitude of the state $|1_{n}\rangle$. Using the discrete
scattering method proposed in Ref.~\cite{ZGLSN08} and according to
the eigenequation $H|\Omega\rangle=\Omega|\Omega\rangle$, we have
\begin{subequations}
\begin{align}
\label{eq:2a}-t(A_{n+1}+A_{n-1})&=(\Omega-\omega)A_{n},\hspace{0.3 cm} n\neq \{l,l+1\},\\
\label{eq:2b}-t'A_{l+1}-tA_{l-1}&=(\Omega-\omega)A_{l},\\
\label{eq:2c}-A_{l+2}-t'A_{l}&=(\Omega-\omega)A_{l+1}.
\end{align}
\end{subequations}

For the coherent transport of a single-photon with the energy
$\Omega=\omega-2t\cos k$, we can assume the following forms for the
probability amplitudes
\begin{subequations}
\begin{align}
\label{eq:3a}A_{n}&=e^{ikn}+re^{-ikn},\hspace{0.3 cm}(n\leq l),\\
\label{eq:3b}A_{n}&=se^{ikn},\hspace{0.3 cm}(n\geq l+1).
\end{align}
\end{subequations}
Here $r$ and $s$ are the reflection and transmission amplitudes,
respectively. Obviously, Eqs.~(\ref{eq:3a}) and (\ref{eq:3b}) are
the solutions of Eq.~(\ref{eq:2a}). Substituting
Eqs.~(\ref{eq:3a}) and (\ref{eq:3b}) into Eqs.~(\ref{eq:2b}) and
(\ref{eq:2c}), we can obtain the transmission coefficient
\begin{eqnarray}
\label{Tcoefficient}T(\lambda,k)=\frac{4(\lambda+1)^{2}\sin^{2}k}{\lambda^{2}(\lambda+2)^{2}+4(\lambda+1)^{2}\sin^{2}k},
\end{eqnarray}
and the reflection coefficient
$R(\lambda,k)=|s|^{2}=1-T(\lambda,k)$. Eq.~(\ref{Tcoefficient})
shows that the reflection and transmission coefficients
$R(\lambda,k)$ and $T(\lambda,k)$ are function of the parameter
$\lambda$ and the wavevector $k$ of the incident photon, and they
are independent of other variables, e.g., the site position
parameter $l$, the cavity frequency $\omega$, and the coupling
constant $t$.
\begin{figure}[tbp]
\includegraphics[width=7 cm,height=5 cm]{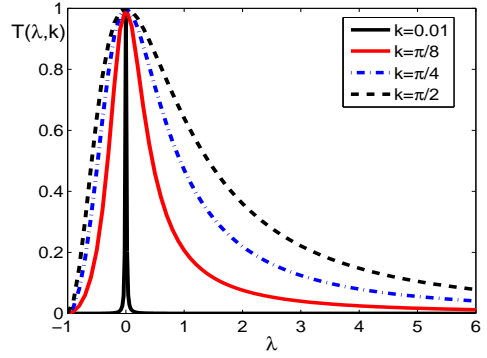}
\caption{(Color online) The transmission coefficient $T$ versus
the parameter $\lambda$ for different wavevectors k=0.01, $\pi/8$,
$\pi/4$, and $\pi/2$ is plotted.}\label{Transmission}
\end{figure}

Equation~(\ref{Tcoefficient}) shows two symmetry relations
$T(\lambda,k)=T(\lambda,-k)$ and
$T(\lambda,\pi/2-k)=T(\lambda,\pi/2+k)$. Therefore we need only to
analyze the transmission coefficient within the region $0\leq
k\leq\pi/2$. In this region, there are four special cases: (1)
$T(\lambda\neq0,0)=0$, when the wavevector $k=0$, for
$\lambda\neq0$, the input single photon is reflected completely; (2)
$T(\lambda=-1,k)=0$, when $\lambda=-1$, the coupling between the
$l$-\textrm{th} and $(l+1)$-\textrm{th} cavities is switched off, so
for any value of the wavevector $k$, the transmission coefficient is
zero; (3) $T(\lambda\rightarrow\infty,k)=0$, when
$\lambda\rightarrow\infty$, namely, $t'\gg t$, the transmission
coefficient is zero for any $k$. Physically, when $t'\gg t$, the
Hamiltonian~(\ref{Hamiltonian}) is approximated to $H(t'\gg
t)\approx -t'(a^{\dagger}_{l}a_{l+1}+a^{\dagger}_{l+1}a_{l})$. The
input photon will stay in the $l$-\textrm{th} and
$(l+1)$-\textrm{th} cavities once it arrives the $l$-\textrm{th}
cavity; (4) $T(\lambda=0,k)=1$, $\lambda=0$ implies $t'=t$, the
present model reduces to the usual bosonic tight binding model, so
the photon with any wavevector can be perfectly transported.

To observe the effect on the transmission coefficient $T$ for
general wavevector $k$ and parameter $\lambda$, in
Fig.~\ref{Transmission}, the transmission coefficient $T$ is
plotted as a function of the parameter $\lambda$ for wavevectors
$k=0.01$, $\pi/8$, $\pi/4$, and $\pi/2$. Fig.~\ref{Transmission}
indicates that there are two regions, $-1\leq\lambda\leq0$ and
$0\leq\lambda$, in which controllable transport of single photon
can be achieved. The transmission coefficient $T$ can be tuned
from $0$ to $1$ by changing the coupling strength $t'$, namely
$\lambda$. When $t'=0$, the transmission coefficient $T=0$. With the
increase of the coupling strength $t'\rightarrow t$, the
transmission coefficient $T$ gradually approaches to 1. For
$t'\geq t$, the transmission coefficient $T$  approaches to $0$
with the increase of the coupling $t'\rightarrow\infty$. In this
region, the larger wavevector $k$ corresponds to the larger
parameter range of $\lambda$. In both regions, the controllable
transport of single-photon with arbitrary wavevector $k$ can be
realized. Therefore, our approach for single-photon transport can
cover complete bandwidth.

\begin{figure}[tbp]
\includegraphics[bb=109 408 429 571, width=7 cm]{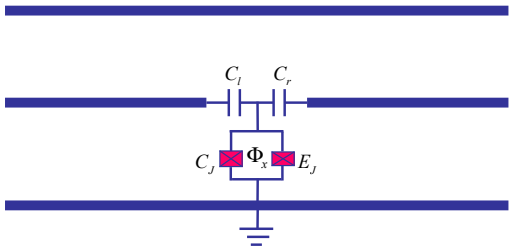}
\caption{(Color online) Schematic diagram for two TLRs (the left
and the right ones), which are coupled to a CPB through two
capacitors $C_{l}$ and $C_{r}$, respectively. The CPB is biased
by a magnetic flux $\Phi_{x}$}\label{TwoTLRs}
\end{figure}

Let us now focus the problem on how to realize controllable coupling
between two TLRs \cite{Switch,miro}. The system we considered is
shown in Fig.~\ref{TwoTLRs}. Two TLRs are coupled to a CPB through
capacitors $C_{l}$ and $C_{r}$, respectively. We assume that the two
TLRs are identical, that is, they have the same length $d$ and
capacitance $C_{0}$ (inductance $L_{0}$) per unit length. We
consider only single-modes of the two TLRs in near resonant with the
CPB. The free Hamiltonian of the two TLRs is
\begin{eqnarray}
\label{TLRhamiltonian}H_{\textrm{TLR}}=\omega
a_{l}^{\dag}a_{l}+\omega a_{r}^{\dag}a_{r},
\end{eqnarray}
where $a^{\dagger}_{l}$ ($a^{\dagger}_{r}$) and $a_{l}$ ($a_{r}$)
are the creation and annihilation operators of the resonant modes
with frequency $\omega$ for the left (right) TLR, respectively.

The CPB is a superconducting loop interrupted by two identical
Josephson junctions with the capacitance $C_{J}$ and the Josephson
energy $E^{(0)}_{J}$. To obtain a tunable Josephson coupling energy,
an external magnetic flux $\Phi_{x}$ is applied through the
superconducting loop. The Hamiltonian of the CPB is
\begin{eqnarray}
H'_{\textrm{CPB}}&=&E_{C}\,n^{2}-E_{J}(\Phi_{x})\cos\varphi,\label{CPBhamiltonian}
\end{eqnarray}
where $n$ is the number operator of Cooper-pair charges on the
island connected to the CPB, and $\varphi$ is the superconducting
phase difference across the Josephson junction. The charging energy
$E_{C}$ and effective Josephson energy $E_{J}(\Phi_{x})$ of the CPB
are $E_{C}=2e^{2}/(C_{l}+C_{r}+2C_{J})$ and
$E_{J}(\Phi_{x})=2E^{(0)}_{J}\cos(\pi\Phi_{x}/\Phi_{0})$,
respectively. Here, we assume that the charging energy and the
effective Josephson energy satisfy the condition $E_{J}(\Phi_{x})\gg
E_{C}$. Under this condition, the spectrum of the lowest energy
levels of the CPB can be described approximately by a harmonic
oscillator~\cite{yingdan2}. That is, we expand
$E_{J}(\Phi_{x})\cos\varphi$ around $\varphi=0$ up to
$\mathcal{O}(\varphi^{2})$, and then Eq.~(\ref{CPBhamiltonian})
becomes
\begin{eqnarray}
\label{e6} H_{\textrm{CPB}}=\omega_{b}b^{\dag}b, \hspace{0.3 cm}
\omega_{b}=\sqrt{2E_{C}E_{J}(\Phi_{x})}. \label{eq:6}
\end{eqnarray}
The annihilation and creation operators $b$ and $b^{\dag}$ in
Eq.~(\ref{eq:6}) are defined in terms of
$\varphi=\sqrt[4]{E_{C}/(2E_{J}(\Phi_{x}))}(b+b^{\dag})$ and
$n=-i\sqrt[4]{E_{J}(\Phi_{x})/(8E_{C})}(b-b^{\dag})$.

We assume that the linear dimension of the CPB is much smaller
than wavelengths of the TLRs, and choose the position of the CPB
at the origin of the axis. Then the quantized voltages at the left
and right TLRs are
\begin{eqnarray}
V_{j}(0)=-i\sqrt{\frac{\omega}{dC_{0}}}(a_{j}-a^{\dag}_{j}),
\hspace{0.3 cm} j=l,r.
\end{eqnarray}
According to circuit theory, we know that the voltage at the island
is $\Phi_{0}\dot{\varphi}/(2\pi)$. Therefore, the Coulomb
interaction induced by the two capacitors $C_{l}$ and $C_{r}$ is
\begin{eqnarray}
\label{Coupling}
H_{I}=\sum_{j=l,r}\frac{C_{j}}{2}\left(V_{j}(0)-\frac{\Phi_{0}}{2\pi}\dot{\varphi}\right)^{2}.
\end{eqnarray}
In fact, capacitors $C_{l}$ and $C_{r}$ induce a direct Coulomb
interaction between the two TLRs with the strength $\propto
C_{l}C_{r}$. However, this direct interaction is much smaller than
the interaction between the two TLRs and the CPB given by
Eq.~(\ref{Coupling}) with strengths $\propto C_{\Sigma l}C_{r}$ and
$\propto C_{\Sigma r}C_{l}$ under the condition $\{C_{\Sigma
l},C_{\Sigma r}\}>>\{C_{l},C_{r}\}$, where $C_{\Sigma
l}=C_{0}d/2+C_{l}$ and $C_{\Sigma r}=C_{0}d/2+C_{r}$ are the sum
capacitors connected to the left and right TLRs,
respectively~\cite{Li}. For instance, using current experimental
parameters~\cite{Frunzio} $C_{0}d/2\sim 1.6$ pF and $C_{l}=C_{r}=6$
fF, we find that the interaction between the TLRs and the CPB is
larger than the direct interaction between two TLRs by three orders
of magnitude.

Using Eqs.~(\ref{TLRhamiltonian}-\ref{Coupling}), the total
Hamiltonian of the system described in Fig.~\ref{TwoTLRs} is
\begin{eqnarray}
\label{e11} H&=&\omega_{l} a_{l}^{\dag}a_{l}+\omega_{r}
a_{r}^{\dag}a_{r}+\omega'_{b}b^{\dag}b\nonumber\\
&&+g_{l}(a_{l}b^{\dag}+ba_{l}^{\dag})+g_{r}(a_{r}b^{\dag}+ba_{r}^{\dag}),
\end{eqnarray}
where we have introduced the renormalized frequencies
\begin{subequations}
\begin{align}
\label{e12a}\omega_{j}&=\omega\left(1+\frac{C_{j}}{dC_{0}}\right),\hspace{1 cm} j=l,r,\\
\label{e12b}\omega'_{b}&=\omega_{b}+(C_{l}+C_{r})\omega_{b}^{2}\left(\frac{\Phi_{0}}{2\pi}\right)^{2}
\left(\frac{E_{C}}{2E_{J}(\Phi_{x})}\right)^{\frac{1}{2}},
\end{align}
\end{subequations}
and the coupling strengths
\begin{eqnarray}
\label{e13}
g_{j}&=-C_{j}\omega_{b}\frac{\Phi_{0}}{2\pi}\sqrt{\frac{\omega}{dC_{0}}}
\left(\frac{E_{C}}{2E_{J}(\Phi_{x})}\right)^{\frac{1}{4}},\hspace{0.3cm}
j=l,r.
\end{eqnarray}
It should be noted that we have made the rotation wave approximation
when Eq.~(\ref{e11}) is derived.

Equation~(\ref{e11}) describes that two TLRs are coupled to the CPB,
which serves as a coupler. To obtain controllable coupling between
the two TLRs, we restrict the system in the large detuning regime,
where the frequency differences between the two TLRs and the CPB are
much larger than their coupling constants, i.e., $\Delta_{l}\gg
g_{l}$ and $\Delta_{r}\gg g_{r}$. Here,
$\Delta_{j}=\omega'_{b}-\omega_{j}$ for $j=l,r$ are the dutuning
between the frequencies of the TLRs and that of the CPB. By
adiabatically eliminating  the degree of freedom of the CPB, we
obtain an effective interaction between the two TLRs. That is, we
perform a unitary transform
$U=\exp[g_{l}(a_{l}b^{\dag}-ba_{l}^{\dag})/\Delta_{l}
+g_{r}(a_{r}b^{\dag}-ba_{r}^{\dag})/\Delta_{r}]$ for the Hamiltonian
in Eq.~(\ref{e11}) and use the Hausdorff expansion up to the first
order in the small parameter $g_{j}/\Delta_{j}$ with $j=l,\,r$, then
we obtain an effective Hamiltonian
\begin{eqnarray}
\label{e15} H_{\textrm{eff}}&=&\omega'_{l}
a_{l}^{\dag}a_{l}+\omega'_{r}
a_{r}^{\dag}a_{r}+g(a_{r}a_{l}^{\dag}+a_{l}a_{r}^{\dag}),\label{eq:15}
\end{eqnarray}
where we have defined the Stark-shifted frequencies
$\omega'_{j}=\omega_{j}+g^{2}_{j}/\Delta_{j}$ for $j=l,r$, and the
effective coupling strength
\begin{eqnarray}
\label{e17}
g=\frac{g_{l}g_{r}(\Delta_{l}+\Delta_{r})}{2\Delta_{l}\Delta_{r}}.
\end{eqnarray}
Note that the effective Hamiltonian of the CPB
$H_{\textrm{CPB}}=\omega''_{b}b^{\dag}b$ with
$\omega''_{b}=\omega'_{b}-g^{2}_{l}/\Delta_{l}-g^{2}_{r}/\Delta_{r}$
has been neglected in Eq.~(\ref{eq:15}). It is obvious that the
Hamiltonian (\ref{e15}) describes an effective interaction between
the two TLRs. According to Eqs. (\ref{eq:6}) and (\ref{e12b}), the
frequency of the CPB can be tuned by the external magnetic flux
$\Phi_{x}$. Correspondingly, the detunings $\Delta_{l}$ and
$\Delta_{r}$ between the TLRs and the CPB can be tuned, thus the
coupling constant $g$ can be tuned. When the detunings are very
larger than the coupling constants between the TLRs and the CPB, the
effective coupling constant $g$ between the two TLRs are negligibly
small, and then the interaction between the two TLRs is switched
off. For example, if we assume that the two transmission line
resonators are identical, i.e., $\Delta_{l}=\Delta_{r}=\Delta$ and
$g_{l}=g_{r}=g'$, and we
 take the parameters~\cite{Frunzio}: $\omega=2\pi\times3$ GHz, $C_{l}=C_{r}=6$ fF, $C_{0}d=1.6$ pF,
$E_{C}=2\pi\times0.35$ GHz, $E^{(0)}_{J}\sim 10^{3}E_{C}$, then we
calculate $g\approx1.1\sim23$ MHz corresponding to
$\cos(\pi\Phi_{x}/\Phi_{0})\approx0.02\sim1$. In this region, the
conditions $E_{J}(\Phi_{x})\gg E_{C}$ and $\Delta\gg g'$ are
satisfied.

In conclusion, we have studied a quantum switch for single-photon
transport in a coupled TLR array with one controllable hopping
interaction. We have found that the controllable single-photon
transport, for an arbitrary wavevector of photons, in the coupled
TLR array can be realized by tuning one of the coupling constants.
How to realize the controllable coupling between two TLRs is also
studied. We have proposed that a CPB serves as a coupler to connect
the two TLRs. In the regime of $E_{J}(\Phi_{x})\gg E_{C}$, the CPB
is approximately described as a harmonic oscillator. Under the large
detuning condition, we have obtained an effective interaction
between the TLRs by adiabatically eliminating the variables of the
CPB. This induced effective coupling can be controlled by the
external magnetic flux $\Phi_{x}$ through the CPB.

This work is supported by the NSFC with Grant Nos. 10474104,
60433050, 10704023, and 10775048; NFRPC Nos. 2006CB921205,
2005CB724508, and 2007CB925204.

\end{document}